\documentclass[aps, prb, superscriptaddress, preprint, showpacs, showkeys,floatfix]{revtex4-1}

\usepackage[cmex10]{amsmath}
\usepackage{amssymb}
\usepackage[english]{babel}
\usepackage{graphicx}
\usepackage[normalem]{ulem}
\usepackage{placeins}
\usepackage{float}
\usepackage[hidelinks]{hyperref}
\usepackage{tabularx,colortbl}
\usepackage{glossaries}
\usepackage{siunitx}
\graphicspath{{imgs/color/}}
\usepackage{xcolor}
\usepackage{filecontents}
\newacronym{TI}{TI}{Topological Insulator}
\newacronym{PTI}{PTI}{Photonic Topological Insulator}
\newacronym{TP}{TP}{Topological Protection}
\newacronym{BPhC}{BPhC}{Bianisotropic Photonic Crystal}
\newacronym{PhC}{PhC}{Photonic Crystal}
\newacronym{TPhC}{TPhC}{Topological Photonic Crystal}
\newacronym{TPHEM}{TPHEM}{Topologically Protected Helical Edge Mode}
\newacronym{TPEM}{TPEM}{Topologically Protected Edge Mode}
\newacronym{TPMW}{TPMW}{Topologically Protected Meta Waveguide}
\newacronym{WG}{WG}{Wave Guide}
\newacronym{QSH}{QSH}{Quantum Spin Hall}
\newacronym{CTPMW}{CTPMW}{Coupled Topologically Protected Meta Waveguides}
\newacronym{PBG}{PBG}{Photonic Band Gap}
\newacronym{PBS}{PBS}{Photonic Band Structure}
\newacronym{Fw}{Fw}{Forward}
\newacronym{Bw}{Bw}{Backward}
\newacronym{CWG}{CWG}{Circular WaveGuide}
\newacronym{TDC}{TDC}{Topological Directional Coupler}
\newacronym{CD}{CD}{Contra-Directional}
\newacronym{LP}{LP}{Linear Polarization}
\newacronym{CP}{CP}{Circular Polarization}
\newacronym{LCP}{LCP}{Left Circular Polarization}
\newacronym{RCP}{RCP}{Right Circular Polarization}
\makeglossaries

\addto\captionsenglish{%
  \renewcommand\figurename{Fig.}
}

\newcommand{\subref}[2]{\figurename{}~\ref{#1}#2}

\begin{document}
\title{Towards Topological Protection based millimetre wave devices}
\author{Gian Guido Gentili}
\affiliation{%
Politecnico di Milano (MI), DEIB (Italy)
}%
\author{Giuseppe Pelosi}%
\affiliation{%
University of Florence (FI), DINFO (Italy)
}%
\author{Francesco S. Piccioli}
\email{piccioli@lens.unifi.it}
\affiliation{%
Politecnico di Milano (MI), DEIB (Italy)
}%
\affiliation{%
LENS and CNR, (FI), INO (Italy)
}%
\author{Stefano Selleri}
\affiliation{%
University of Florence (FI), DINFO (Italy)
}%

\begin{abstract}
Feasibility of Topological Metawaveguides supporting helical propagation in the microwave range has been recently proven. The advantages of unidirectional propagation supported by such waveguides however can only be exploited in real devices if topological modes are endowed with the capability to interact within themselves as well as with trivial modes.
Here we show a modal launcher to interface a topological metawaveguide with conventional circular waveguides with negligible reflection and we exploit the properties of coupled topological modes to show a proof of concept of a topological contra-directional coupler.
\end{abstract}

\maketitle

\glsresetall
In recent years a new field of physics, known as \textit{topological photonics}, has rapidly emerged \cite{wang2008,wang2009,haldane2008,khanikaev2012,rechtsman2013, lu2014,ma2015,lai2016,rechtsman2016, Khanikaev2017,ozawa2019}. Although its initial aim was to emulate, employing higly controllable photonic systems, topological effects originally discovered in quantum matter\cite{Kane2006,Barnevig2006, Hasan2010}, soon enough \glspl{PTI} appeared as an exciting platform for the realization of new robust and low-loss photonic devices. As for condensed matter, \glspl{PTI} are \textit{insulating} in the sense that they have a complete \gls{PBG} inhibiting traveling bulk photonic states. However their topologically non-trivial order, marked by indexes as Chern or winding numbers\cite{rudner2013,nathan2015}, endows their edges with fascinating properties. Although these edge properties emulates their condensed matter counterparts, the flexibility in the design of artificial photonic media\cite{joannopoulos,longhi2009} also allows to observe phenomena that cannot be easily observed or do not at all have solid-state analogues\cite{ma2017, quelle2017}.\par
Possibly the most important property of \gls{PTI} edges is that interfaces between two \glspl{PTI}, or between a \gls{PTI} and a non-topological photonic insulator, support gapless unidirectional modes across the common \gls{PBG}, also called \glspl{TPEM}. Because of the waveguiding features being related to the topology of the confining mirrors, these interfaces are often named \glspl{TPMW}. Whereas topological characteristics are invariant under omeomorphic transformations, \glspl{TPMW} have remarkably robust waveguiding properties even against imperfections of the confining \glspl{PTI} \cite{wang2008,wang2009}. The sinergy between their robustness and the exceptional feature of unidirectionality makes them promising for a broad range of applications in integrated photonics and nano optics \cite{ma2015,lai2016}. Reflection-less and unidirectional propagation around disordered regions \cite{khanikaev2012}, sharp bends \cite{ma2015}, and large defects \cite{yves_2017} has been shown both theoretically and experimentally employing different kind of \glspl{TPMW}.\par
However, when it comes to real devices, the thrilling robustness of topological modes is rather quickly converted into a double-edged weapon. Indeed even simple applications, as energy conveyance, require some degree of interaction, for instance with a source and a detector, yet topological modes hardly interact with anything else. Therefore, if topological propagation has to be exploited in real world devices, there are two gaps to overcome. On one side one needs to efficiently convert a non-topological mode into a topological one. This includes both being able to excite a topological mode with a non-topological source, and extract power flowing in a topological mode to detect it with a conventional detector, both with minimum losses in the process. On the other side an all-topological platform for signal processing is desirable, but a fundamental step towards its implementation is the study of interactions between different topological modes.\par
In this contribution we address both the aforementioned problems using a similar approach based on local breaking of \gls{TP}. First we are able to observe an excellent transition between a conventional mode and a topological one by carefully designing a transition region and later applying optimization methods derived from microwave engineering. The designed interface can be used both for an efficient excitation of topological modes, for which we observe a reflection coefficient as low as -10db over the whole bandwidth, and for a full-vectorial detection and characterization of power flowing through a topological channel employing, for instance, S-parameters which are directly measurable from commercial VNAs. On the other hand we qualitatively study the interaction between counterpropagating topological modes. We observe that a local breaking of topological protection results in non-null coupling between counter propagating topological modes which can be used to realize directional and contra-directional couplers. Finally we show a proof of concept for such a \textit{Topological Directional Coupler}
\subsection*{Topological Properties}
\glspl{PTI} have been realized with a large number of micro and nanophotonic platforms. As a general classification one can define two types of systems. First, topological systems supporting chiral unidirectional edge modes, characterized by a $\mathbb{Z}$ topological invariant that counts the number of unidirectional modes across a given bandgap. These systems can be realized by explicitly breaking time-reversal symmetry either using ferromagnetic materials\cite{wang2008, wang2009} or exploiting Floquet physics\cite{rechtsman2013,maczewsky2017,mukherijee2017}. Second, topological systems supporting uncoupled counter-propagating topological modes, characterized by a $\mathbb{Z+Z}$ topological invariant and that can be realized also in presence of time-reversal symmetry\cite{khanikaev2012,ma2015,ma2016,barik2016}. Very recent researches also demonstrated $\mathbb{Z}_2$ topological systems in which Floquet temporal modulation enables time-reversal symmetric systems with counter-propagating modes persisting even after inter-spin coupling\cite{maczewskyz2}. However uncoupled counter propagating modes are attractive \textit{per se} since reflectionless propagation can be observed without any magnetic bias or temporal modulation, yet light propagation direction is usually entangled with some additional property of the EM fied (such as polarization) which is interesting for applications as polarization discriminators. Among all existing proposals we base our results upon the bianisotropic metawaveguide concept theoretically introduced in and experimentally demonstrated in \cite{lai2016}, because of its convenient operating frequency range and relatively easy implementation.
\par
\begin{figure}[t]
\begin{center}
\noindent
	\includegraphics[width=\columnwidth]{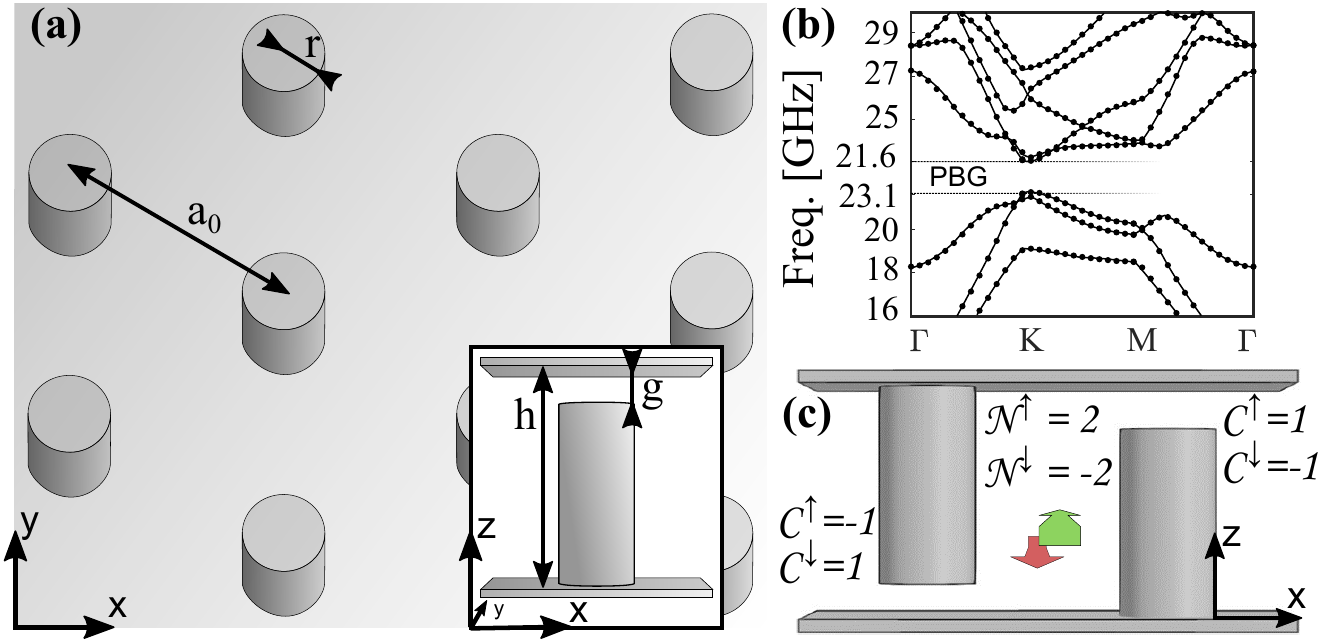}
  	\caption{
  	\textbf{(a)} Schematic of \gls{TPhC}, $a_0 = 10mm$, $r = 1.725a_0$, $g = 0.15a_0$, $h = a_0$.
  	\textbf{(b)} \acrfull{PBS} of the \gls{TPhC} with the complete \gls{PBG} highlighted. 
  	\textbf{(c)} Topological interface between two $z$-symmetry reversed \glspl{TPhC}. The number of unidirectional edge modes for every spin state is given by the difference between the confining spin-Chern Numbers.
  	}
  	\label{imgTPMW}
\end{center}
\end{figure}
The \gls{PTI} (shown in \subref{imgTPMW}{a}) consists in a triangular arrangement of metallic rods, asymmetrically perturbed along the $z$ direction as to introduce an air-gap between the bed-of-rods and one of the confining metallic planes. The eigenfrequencies of the \gls{PTI} (reported in \subref{imgTPMW}{b}) have a complete \gls{PBG} of $\approx 1.5\mathrm{GHz}$ centered around $22.33\mathrm{GHz}$. Considering \gls{CP} basis to express the fields, the four modes, two on the upper and two on the lower edge of the \gls{PBG}, can be expressed in terms of two uncoupled set of two modes each, one with \gls{RCP} and one with \gls{LCP}. Because the two sets are uncoupled one can calculate the Chern number of both sets asunder, denoting them as as $\mathcal{C}^\uparrow$ and $\mathcal{C}^\downarrow$. While time reversal symmetry enforces the total Chern number $\mathcal{C} = \mathcal{C}^\uparrow + \mathcal{C}^\downarrow$ to be zero, the spin-Chern numbers $\mathcal{C}^{\uparrow/\downarrow}$ might individually acquire opposite non-zero integer values. In such case the \gls{PTI} is equivalent to two uncoupled set of $\mathbb{Z}$-type \glspl{PTI} with opposite chirality. For the system under study Ma \textit{et al.}\cite{ma2017} calculated $\mathcal{C}^{\uparrow/\downarrow} = \pm 1$, with the spin-Chern numbers changing sign also as a result of a z-inversion (relocation of the air gap from the top to the bottom edge). As a matter of fact, two copies of the \gls{PTI} with reversed position of the air gap, placed one close to the other in such a way that the hexagonal symmetry still holds globally, give rise to a topological domain wall across which the spin-Chern invariants of the structure changes from $\pm 1$ to $\mp 1$.
An interface between media with different topological invariants supports a number of edge modes that is equal to the difference between the topological indexes (a fundamental principle known as Bulk-Edge Correspondance). Thereof two $\bf \psi^\uparrow$ unidirectional modes are expected as a result of the $\mathcal{C}^\uparrow$ difference as well as two $\bf \psi^\downarrow$ modes associated to the $\mathcal{C}^\downarrow$ difference. These two sets of modes are referred to as \textit{quasi-spin} modes\cite{ma2015, ma2016, barik2016, he2016}; they carry energy in opposite directions and, as long as topological protection is maintained, they cannot scatter one into the other. Since the propagation direction of such topological modes is locked with their polarization state it is also common to name them \gls{TPHEM}. We note here that $\mathbb{Z}+\mathbb{Z}$ topological systems supporting \gls{TPHEM} are different from fermionic systems with Spin-Orbit coupling exhibiting $\mathbb{Z}_2$ topological insulating phase. In the latter case gap-less topological edge modes are preserved even in the presence of inter-spin coupling (Rashba coupling) while in $\mathbb{Z+Z}$ \glspl{PTI} topological protection has to be assisted by additional symmetries. However, as it will be clear in the following, the somehow reduced protection of $\mathbb{Z+Z}$ insulators is an enabling feature to obtain the exotic coupling features that will be described in the second section of this work.
\begin{figure}[t]
\begin{center}
\noindent
	\includegraphics[width=\columnwidth]{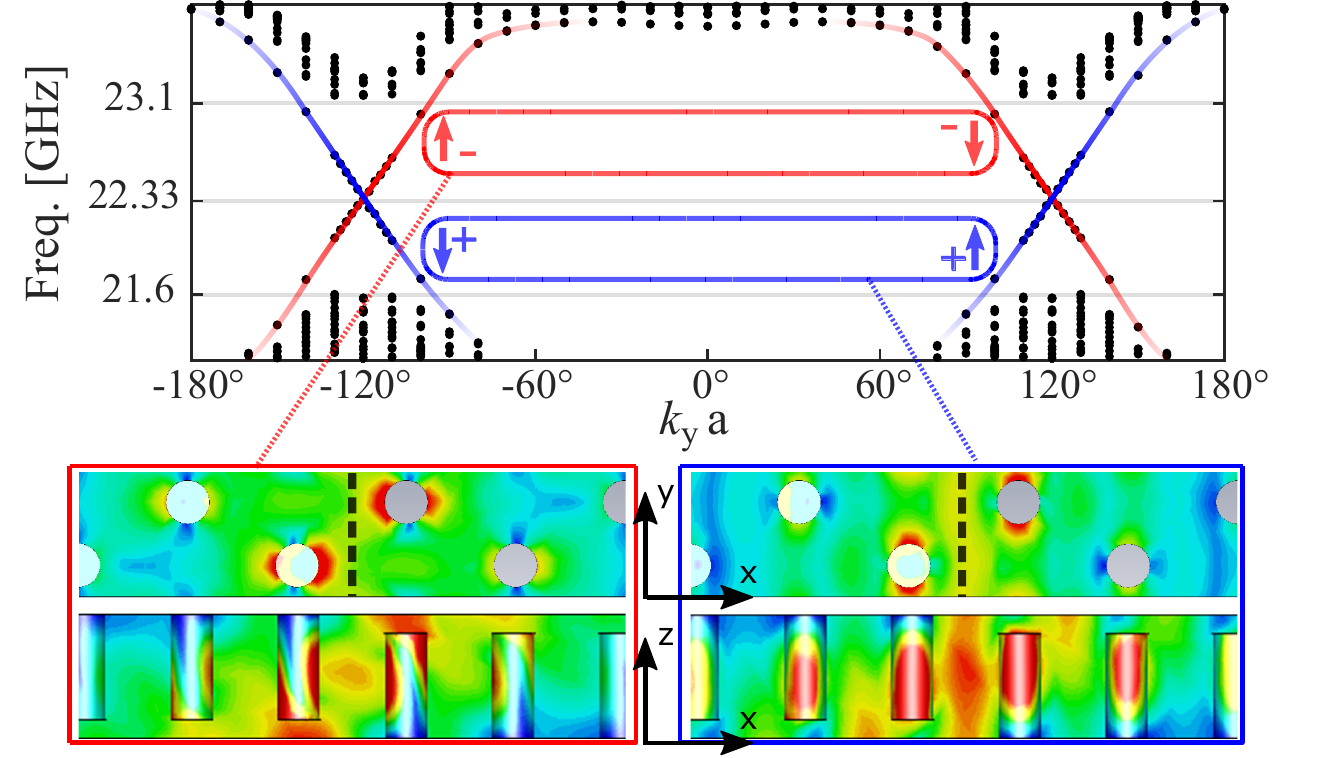}
  	\caption{
    (color online) \textbf{Top:} \gls{PBS} of the topological interface. Arrows are spin states and signs are modal effective index sign. \textbf{Bottom}: Electric field amplitude in the longitudinal and transverse direction for the \glspl{TPHEM}. The spin state is determined by the time evolution of the electric field in the air gap region. 
  	}
  	\label{imgMODES}
\end{center}
\end{figure}
\section{Circular Waveguide Launcher}
The propagating mode of the structure within a given bandwidth can be excited by an antenna inserted in the \gls{TPMW} itself. If an antenna with \gls{LP} is used, the whole set of propagating modes are simultaneously excited leading to four modes, bidirectional, propagation. A more interesting situation is that of an antenna that matches the specific time evolution of the only forward (backward) modes; in the latter case only one kind of pseudo-spin will couple to the excitation, resulting in dual mode,unidirectional propagation. In the initial theoretical proposal \cite{ma2015} a source with \gls{LCP}, rotating in the plane containing the structure, has been used to simulate unidirectional excitation. However point dipoles are ideal sources and, while acceptable approximations of ideal dipoles can be built, it is unpractical, if not nearly impossible, to build such sources inside the structure. Indeed in the first experimental work \cite{lai2016} a short dipole antenna has been used for a broadband excitation of the structure by inserting it into the \gls{TPMW} through a small hole. Although a short dipole is a more practical antenna it is not circularly polarized, therefore both Forward and Backward modes are excited with this scheme. We also note that previous attempts to excite topological modes have not considered the characteristic impedance of the topological modes. Indeed if the antenna is not properly matched to the mode's impedance only a fraction of the feeding power will couple to the travelling mode, being the most part of it reflected towards the power source. If in early experiments this does not represent an issue, it becomes of the uttermost importance if \glspl{TPMW} shall be used as a component in real life devices.\par
In this section we propose a design for a modal launcher based on a \gls{CWG}. We show how impedance matching can be effectively used to optimize such transition over a relatively large bandwidth thus obtaining low loss excitation of helical modes both for injection and extraction of a test signal. In spite of its simplicity, our approach is easily generalized for any antenna geometry, such as planar or slot antennas \cite{pierslauncher}. The low-loss transitions obtained with established optimization methods from microwave engineering provide a way to fully characterize transmission of any kind of \gls{TP} based microwave device.\par
\paragraph{Design of the launcher:}
\begin{figure}
\begin{center}
\includegraphics[width = \columnwidth]{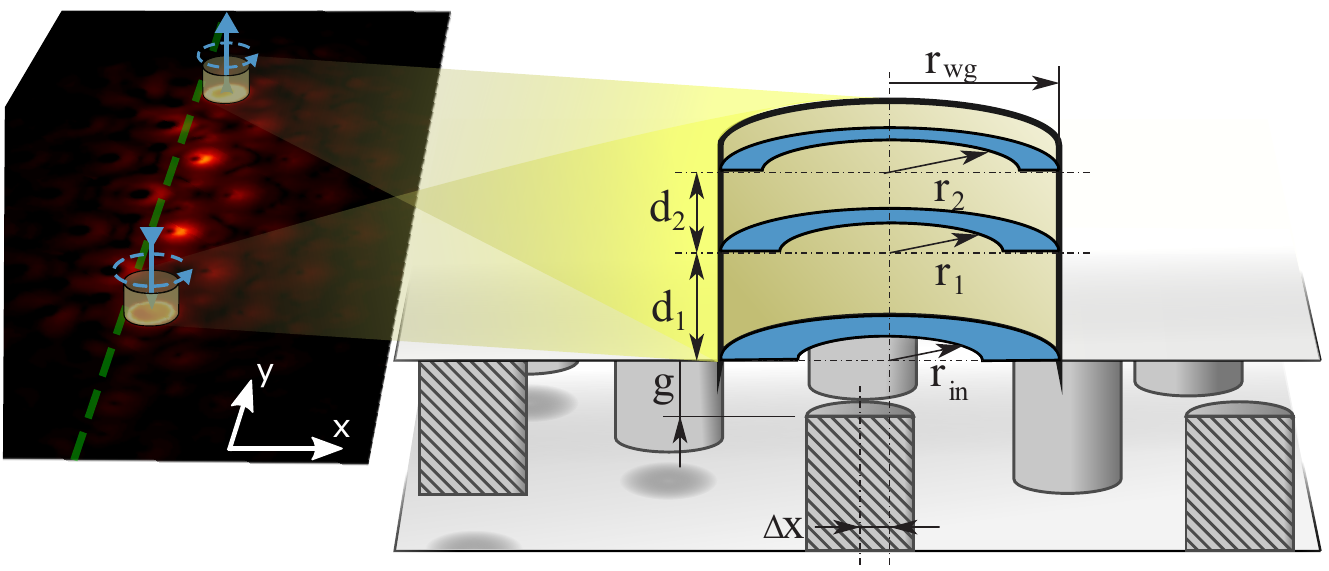}
\caption{(color online)
\textbf{left:} Absolute value of the Poynting vector for a \gls{TPHEM} propagating from a \gls{CWG} launcher to a second one placed at a distance of $9a_0$. \textbf{right:} Schematic of the \gls{CWG} launcher.}
\label{imgCWGa}
\end{center}
\end{figure}
Our design is based on the degenerate ${\rm TE}_{11}^{x,y}$ modes supported by a \gls{CWG}. These orthogonal, \gls {LP}, modes can be easily transformed into a couple of \gls{CP} clockwise and anti-clockwise modes by a transformation matrix.
\begin{equation}\label{cptransformation} 
{\bf T} = \frac{1}{\sqrt{2}}
\begin{bmatrix} 1 &j \\
1 &-j \\ 
\end{bmatrix}
\end{equation} 
A circular hole cut in the top metal plate of the \gls{TPMW} can be used to couple a -$z$ directed \gls{CWG} to the topological structure with an excellent matching of the \glspl{TPHEM}'s pseudo-spin. The radius of the \gls{CWG} is a fixed design constant and is chosen in order to have only first order modes propagating. After fixing the \gls{CWG} radius the design of the modal launcher is carried out in two distinct steps. At first we consider the transition itself, whose frequency response strongly depends on the coupling hole position and radius and subsequently we optimize the coupling with the \gls{CWG} in order to minimize reflections.\par
\begin{figure}
\begin{center}
\includegraphics[width = \columnwidth]{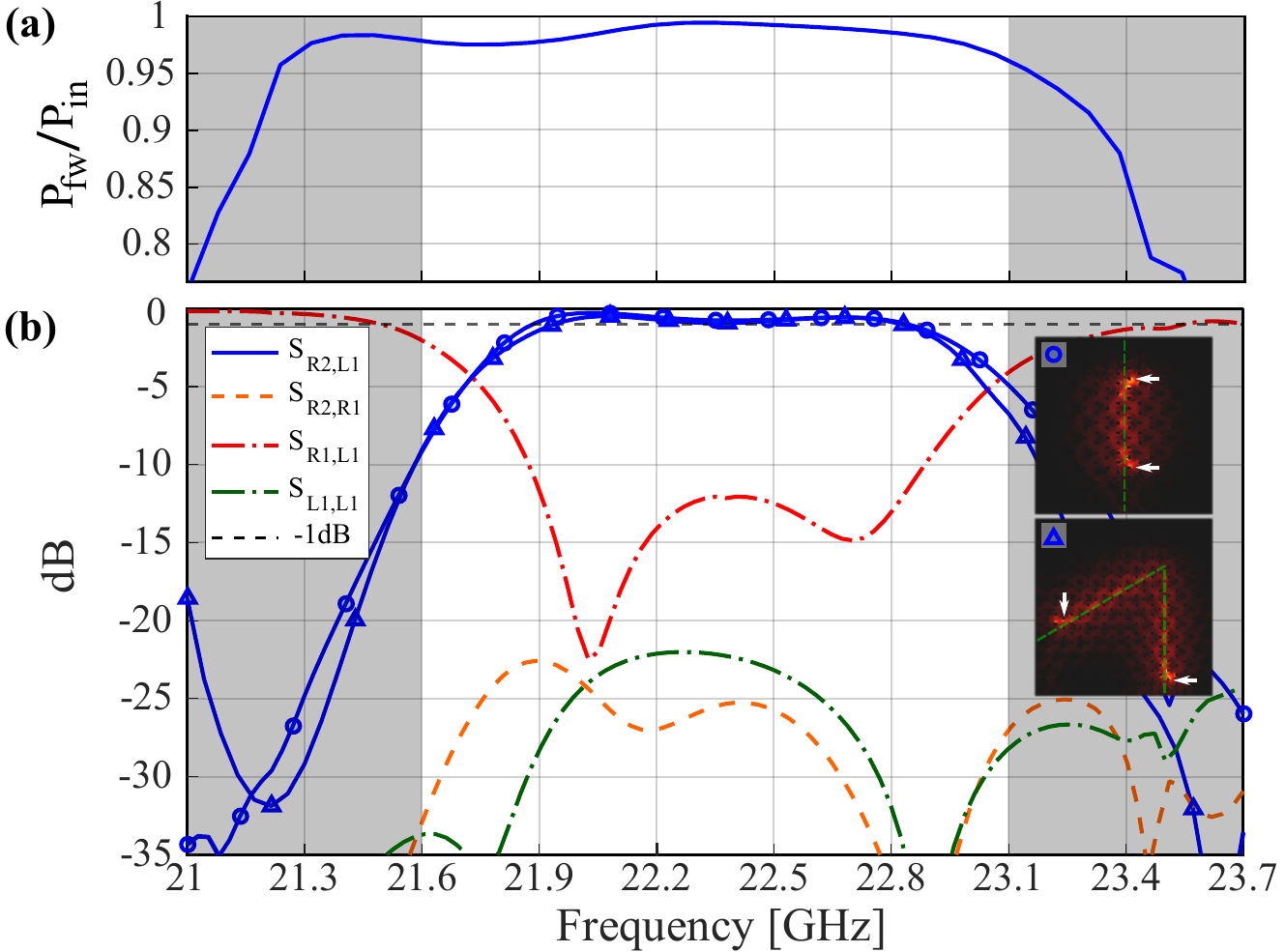}
\caption{(color online)
\textbf{(a) } Ratio between \gls{Fw} and total outgoing power as a function of excitation frequency.
\textbf{(b) } \textit{Solid lines:} \gls{LCP}-to-\gls{RCP} transmission of the straight (circles) and bent (triangles) \glspl{TPMW}. \textit{Dash-point:} Single port co and cross polarization reflection coefficients. \textit{Short dash:} \gls{RCP}-to-\gls{RCP} transmission of the straight \gls{TPMW}. \textit{Inset:} Poynting vector at $z = h/2$; \gls{TPMW} is indicated with green short-dash line while white arrows indicate \gls{CWG} launchers positions.}
\label{imgCWGb}
\end{center}
\end{figure}

 
Due to the underlying hexagonal symmetry of the lattice we expect an asymmetric behaviour of the transition with respect to the two \gls{LP} modes of the \gls{CWG}. This might give rise to a cross-polarization term across the transition which causes an incident L(R)CP waveguide mode to excite a fraction of R(L)CP mode in the \gls{TPMW}. Eventually this will result to radiation in the undesired direction and unwanted losses; the goal of the first step is minimizing this source of losses. We also observe that purest \gls{CP} of the topological mode is located in a point $\Delta x \approx 0.65r$ away from the interface starting from the rod's center. Using that point as the excitation axis, we vary the hole radius and calculate the pointing vector flux across both ends of the waveguide, $P_{\rm fw}$ and $P_{\rm bw}$. Having defined the ratio $\eta = P_{\rm fw}/(P_{\rm fw} + P_{\rm bw})$ as a figure of merit for the effectiveness of the \gls{Fw} excitation, we find a maximum of $\eta$ for $r_{\rm in}/r = 2.25$ (see \subref{imgCWGb}{a}).\par
To optimize the coupling to the \gls{CWG} we first obtain the frequency dependent S-parameters matrix relative to the \gls{LP} waveguide modes (${\bf S}^L$) from the de-embedded input impedances of the optimized window. A congruent transformation can be used to transform the S-parameters to a basis of \gls{CP} modes \cite{graves1956, gentili2018}
\begin{equation}\label{congruenttransf} 
{\bf S}^C = {\bf T}^*{\bf S}^L {\bf T}^{\dag}
\end{equation}
Where the $\bf T$ matrix is given in \eqref{cptransformation}. Note that the simple conjugate in the first term of the RHS in \eqref{congruenttransf} inverts the rotation direction of R(L)CP modes for the reflected (outgoing) waves and is required in order to maintain the symmetry of the transformed ${\bf S}^C$ matrix \cite{gentili2018}. The diagonal terms of the ${\bf S}^C$ matrix, related to a cross-polarization reflection and given by $S_{l(r),l(r)} = (S^L_{x,x} - S^L_{y,y})/2 \pm iS^L_{x,y} $ can be used as a first approximation for evaluating the asymmetry of the transition. For the optimized window this term is smaller than 20dB over the entire topological bandwidth, this confirms that the S-parameters for the orthogonal \gls{LP} modes are approximately equal to each other and that the \gls{CWG} coupling can be optimized using axisymmetric elements, having the same effect on both \gls{LP} modes. The cascade of a \gls{CWG} with the above mentioned window can be represented as a transmission line with a characteristic impedance given by the generalized impedance of the first order \gls{CWG} modes, connected to a frequency-dependent load with impedance given by the input impedances window itself. In such a configuration an incident wave will exhibit a reflection on the load given by \eqref{Reflection}
\begin{equation}\label{Reflection}
\Gamma  = \frac{Z_L-Z_0}{Z_L+Z_0}
\end{equation}
Where $Z_{L}$ is the load impedance and $Z_{0}$ is the transmission line characteristic impedance. Load matching is the procedure of using a matching network, placed between the load and the transmission line, in order to modify the equivalent load impedance with the goal of matching the transmission line's characteristic impedance and eliminate reflections \cite{Selleri2011}. For narrow band operation the procedure is easily performed in a deterministic way by first moving, along the transmission line, to a distance from the load in which its normalized impedance $Z_L/Z_0$ has unitary real part, and then eliminating the residual reactance by placing an element with purely imaginary opposite reactance in parallel to the load. Broadband operation, conversely, involves multi-stage matching with a high number of degrees of freedom which generally requires a numerical optimization strategy. In the present case, inductive irises can be used for realizing purely imaginary loads. The degrees of freedom for the design of a double irises matching circuit are illustrated in \subref{imgCWGa}. In addition, the irises' thickness has also been parametrized in order to fine tune the matching network response.\par
The optimization is performed with a genetic algorithm\cite{agastra2014, agastra2008} in which the response of the matching circuit is simulated with the Mode Matching method \cite{Itoh}.
In the end of the optimization routine we obtain a reflection coefficient lower than -10dB over a bandwidth of 1.1GHz with peaks of -20dB, using irises with thickness $t = 0.5$mm, distances $d_1 = 6.75$mm, $d_2 = 7.65$mm and radii $r_1 = 0.62r_{\rm wg}$, $r_2 = 0.66r_{\rm wg}$(see \subref{imgCWGb}{b}). \par
The matching bandwidth, calculated as the spectral region with reflection coefficient lower than -10dB, covers $ \approx 73\% $ of the bulk \gls{PBG}. That is sufficient to characterize with S-parameters typical features of topological propagation, such as reflection-less propagation of polarization-locked waves around sharp bends across all the bulk \gls{PBG}.
\paragraph{Probing topological protection:}
As previously said, rotating sources matched to the quasi-spin temporal evolution as the one considered in this and previous works\cite{ma2015, sandiego2017}, can be employed to select a specific quasi-spin degree of freedom but not to excite a single propagating mode. Indeed both positive and negative effective indexes \gls{Fw} modes are excited in response to a \gls{LCP} input, with arbitrary amplitude-phase relation that is typical for every specific launcher design. Since the effective load of the \gls{TPMW} is strongly dependent on the excited fields, the transition will behave as expected only at those points in which the amplitude-phase relation of the propagating modes is equal to the one at the excitation point. These points of the \gls{TPMW} are the only ones in which an output interface is able to efficiently convert a \gls{TPHEM} to a conventional waveguide mode and are identified as \textit{extraction points}. The distance between an excitation point and an extraction point, or between two possible extraction points is given by the beating length of the two propagating modes, which is in turn determined by the difference in the \gls{TPHEM} propagation constants, $\Delta_{\pm} = k_+ - k_- = 4\pi/3a_0$. Luckily enough $\Delta_{\pm}$ turns out to be approximately constant across all the \gls{PBG} so that the interference period $P_y$ along the propagation direction is also constant and can be calculated as the (integer) number of reticular constants required to obtain a phase difference of $\Delta\phi=2\ell\pi = \Delta_{\pm}P_y$. Since $P_y/a_0$ must be an integer, setting $\ell=2$ one obtains $P_y=4\pi/\Delta_{\pm} = 3a_0$. This super-reticular periodicity can be observed in \subref{imgCWGa}{} and it enforces the distance $L_{i/o}$ between an input and an output port to always be $L_{i/o} = mP_y = 3 m a_0$, with $m$ integer. \par
In order to demonstrate the effectiveness of the proposed launcher in characterizing topological propagation we model a straight \gls{TPMW} and a bent one, with a very sharp $\deg{120}$ turn. We place input and output interfaces ad appropriate distances and measure the scattering parameters between the two ports. While an \gls{RCP} input is attenuated  more than 20 dB before exiting from the output port, an \gls{LCP} input is transmitted with maximum total losses of 1dB over a fractional bandwidth of 4.4\% which represents the 64\% of the \gls{PBG}; transmission with maximum total losses of 3dB is instead observed over 87\% of the \gls{PBG}. Moreover, the transmission spectra of the straight and bent \glspl{TPMW} are nearly equal inside the matching bandwidth, which confirms the topological nature of \glspl{TPHEM} propagation.\par
As a final note we stress out that input and output coupling happens through out-of plane propagation, and thus propagation direction is inverted for an input and an output wave. Although an input \gls{LCP} mode on port 1, rotating in the counter-clockwise direction, is coupled to $\Psi^\uparrow$ modes flowing in the \gls{Fw} direction, the corresponding transmitted mode in the output port 2 is still rotating in the counter-clockwise direction but is defined, accordingly to \eqref{congruenttransf}, as an outgoing \gls{RCP}. Only with these definitions is the reciprocity of the structure conserved since an incoming \gls{RCP} wave on port 2 is now reciprocally coupled to a $\Psi^\downarrow$ \gls{Bw} mode and transmitted to the \gls{LCP} output at port 1.
\section{Coupled Topological Modes}
In the second part of this paper we focus on the interaction between different topological modes, coupling through evanescent fields. Topological modes are not expected to show strong evanescent coupling, indeed one of the requirements for topological protection is the complete absence of coupling between different modes. However perturbations play a huge role in this case; it is possible to devise regions in which topological protection is broken, and use these regions to obtain some degree of interaction between topological modes. We study in a qualitative way the interactions between closely placed bianisotropic \gls{TPMW}, in a structure that we call \gls{CTPMW}. We show that they exhibit peculiar coupling effects with a strong spectral dependence which we explain as the interplay of two different coupling phenomena namely \textit{spin} and \textit{inter-spin} coupling. Eventually we summarize our findings by illustrating a proof-of-concept of a directional coupler for topological states.
\subsection*{Dual symmetric interface}
\begin{figure}[t]
\begin{center}
\noindent
\includegraphics[width=\columnwidth]{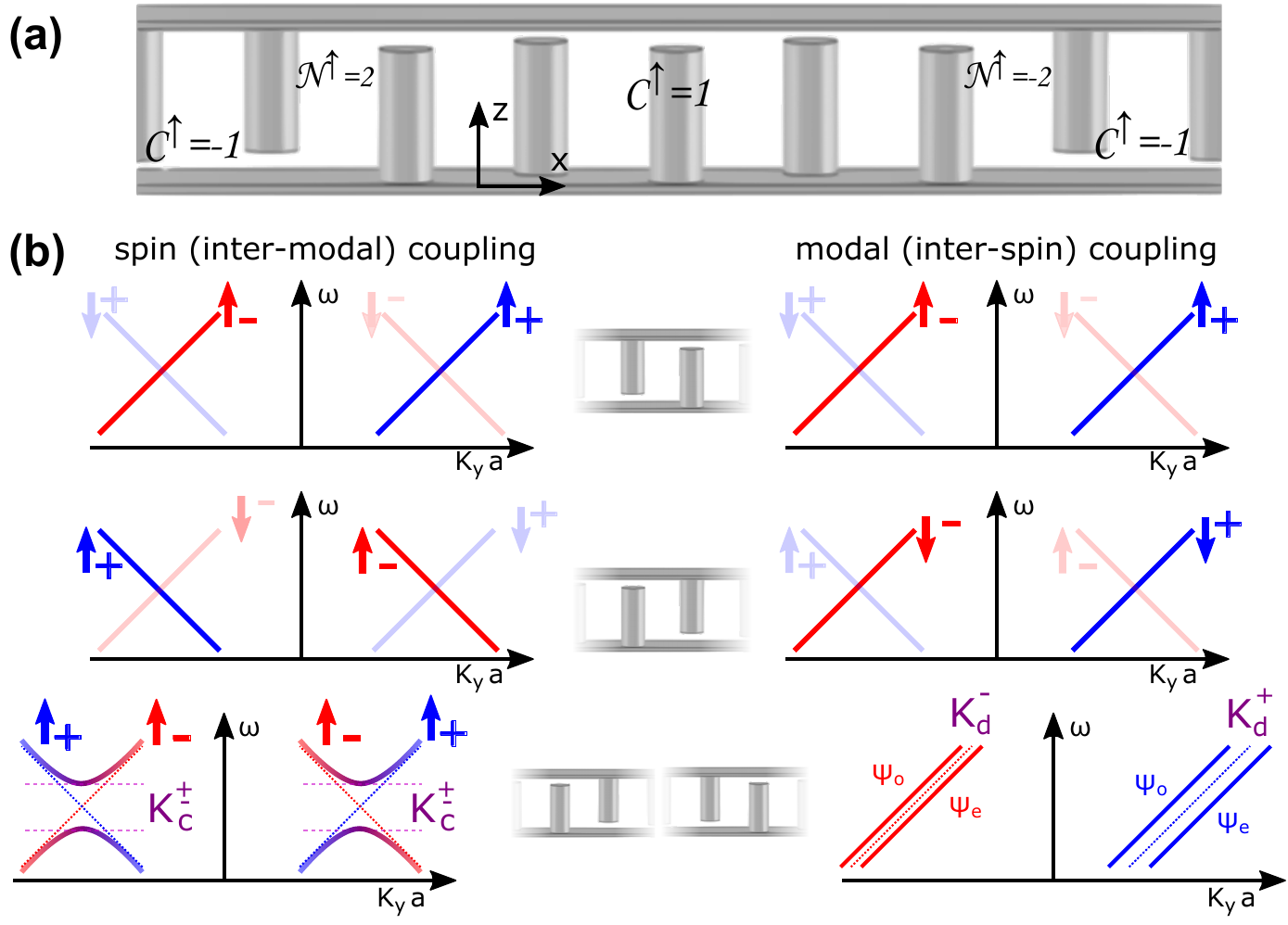}
\caption{(color online)
\textbf{(a)} Double topological interface with $N=5$ interstitial rods. \textit{Spin up} modes have different propagation directions in the two \glspl{TPMW}
  	\textbf{(b)} \textit{left:} $\Psi^\uparrow$ eigenmodes of the left interface couple to $\Psi^\uparrow$ eigenmodes of the right interface when phased matched, causing anti-crossing. \textit{right:} $\Psi^\uparrow$ modes of the left interface couple to $\Psi^\downarrow$ modes of the right interface across all the \gls{PBG} causing symmetric/anti-symmetric pair splitting.
}
\label{imgCTPMW}
\end{center}
\end{figure}
Figure \ref{imgCTPMW}a shows the cross section of a dual symmetric interface that can be obtained by sandwiching a number $N_s$ of up-facing rods between two bulk crystals of down-facing rods. Recalling that $\mathcal{C}^{\uparrow/\downarrow}$ reverses its sign at any relocation of the air gap, because of the Bulk-Edge Correspondence principle the two \glspl{TPMW} needs to have inverted handedness. Moreover the \gls{CTPMW} structure is symmetric along the $x$ axis as opposed to the uncoupled \gls{TPMW}; as a consequence the modes of the right interface in \subref{imgCTPMW}{a} are exactly the same modes of the left interface, apart from a change in the sign of the wave vector $k_y$ and propagation direction. A complete and rigorous coupled mode formulation for the problem would require to take the interplay between all four modes of the two \glspl{TPMW} into account, leading to 16 coupled mode equations. However, the problem can be dramatically simplified by neglecting couplings between different modes of the same \gls{TPMW} because of their orthogonality in the uncoupled case, and dividing the inter-\gls{TPMW} couplings into only two distinct phenomena: \textit{Spin} couplings and \textit{inter-spin} couplings. These two phenomena are schematically depicted in the left and right sides respectively of \subref{imgCTPMW}{b} and will be illustrated in the following. 
\paragraph{Spin (inter-modal) coupling: }
The dispersion relation of an uncoupled \gls{TPMW} (\subref{imgMODES}{}) shows two degeneracy points ($k=\pm\ang{120}$ at $f=22.33$GHz) between $\Psi^\uparrow$ and $\Psi^\downarrow$ modes. These modal intersections are protected by spin-orthogonality condition meaning that counter-propagating modes belong to uncoupled spin subspaces thus cannot give rise to anticrossing. In the dual symmetric case, however, counter-propagating modes of different waveguides belongs to the same spin subspace (because of the inverted guides handedness) and coupling is not prohibited. Since spin-coupling involves interaction between $n_{\rm eff}^\pm$  and $n_{\rm eff}^\mp$ modes, but with the same spin, it can also be referred to as an \textit{inter-modal} coupling; it is mediated by the phase matching condition and as such will be present only in a small frequency range, ultimately resulting in avoided crossing that opens a small gap in the interfaces modes' dispersion. 
A straightforward consequence of the described spin coupling mechanism is that a \gls{Fw} mode traveling in one of the two \glspl{CTPMW} will progressively leak its eneregy to a \gls{Bw} mode of the other \gls{TPMW}, a phenomenon known as \acrlong{CD} coupling \cite{sandiego2017}. While in conventional \gls{CD} couplers the required phase matching between \gls{Fw} and \gls{Bw} modes is satisfied by an appropriately designed Bragg grating between the two waveguides, it is automatically present in our \glspl{CTPMW} structure because of the modes symmetries. As it will be further discussed in the following, Topological \gls{CD} coupling offers several advantages with respect to  conventional Bragg-assisted one. 
\paragraph{Inter-spin (modal) coupling} Away from the degeneration frequency spin-coupling cannot happen because of phase matching not being satisfied. At the same time spin-orthogonality seemingly prevents coupling of modes with opposite spin thus preventing any kind of coupling in the \gls{CTPMW} structure. However, topological order is partially lost in the central region because of the mutual perturbation between the two \glspl{TPMW} and the finite size of the central domain. This breaks the orthogonality between spin-reversed (and co-directional) modes of the two waveguides, which can interact. In this regime the coupling is weak, even if the \glspl{CTPMW} are close, because power transfer from one \gls{TPMW} to the other involves a change of spin (thus polarization). The coupling length $L_0$ is expected to be several reticular constants long and, in principle, different between positive and negative index modes because of their different degrees of edge localization.\par
\begin{figure}
\begin{center}
\noindent
	\includegraphics[width=\columnwidth]{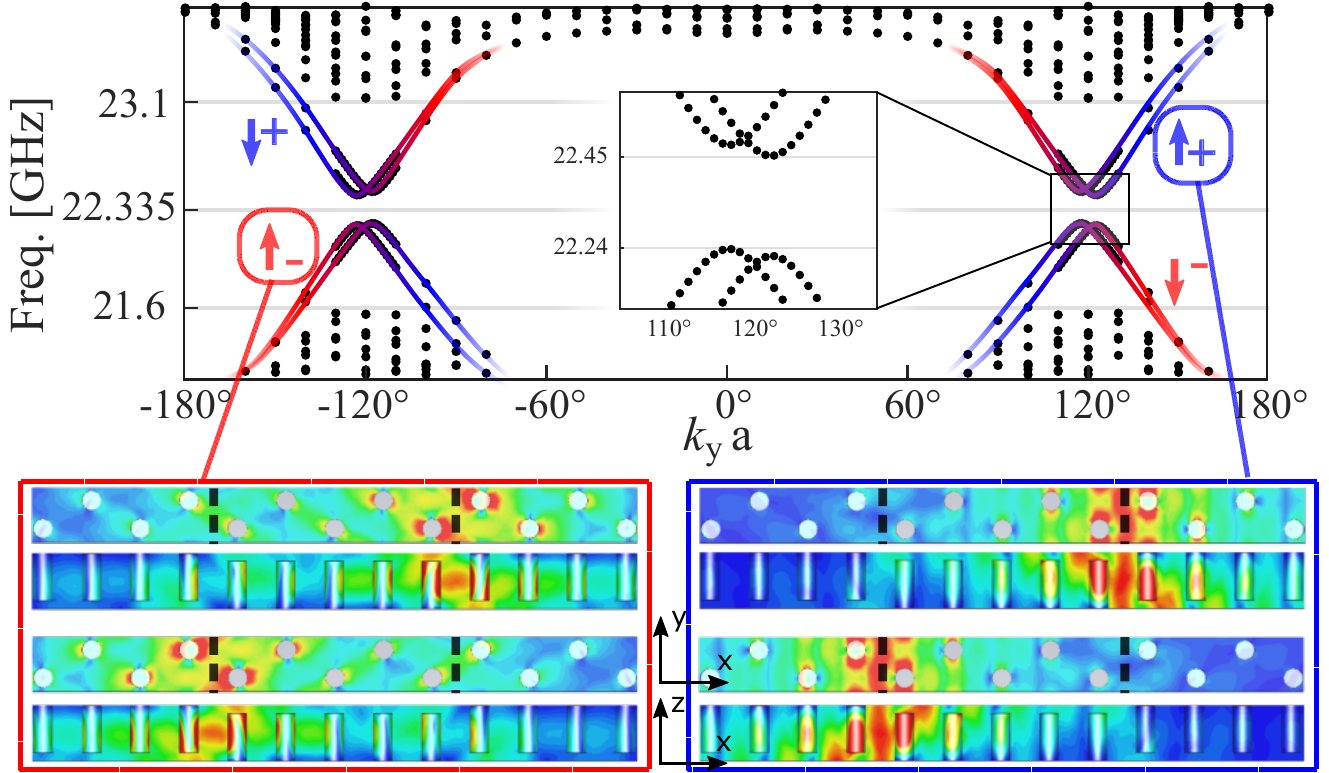}
  	\caption{(color online)
  	\gls{PBS} of the double interface. \textbf{Inset}: longitudinal and transverse electric field amplitudes for evenly (top) and oddly (bottom) combined eigenmodes couples. 
  	}
  	\label{imgCMODES}
\end{center}
\end{figure}

The simultaneous effects of both coupling mechanisms can be observed looking at the modes of an infinite strip comprising both \glspl{CTPMW} (\subref{imgCMODES}{}). Every mode of the uncoupled \gls{TPMW} divides into a couple of symmetric/antisymmetric modes (also called super-modes), confirming that there is directional coupling between the two \glspl{CTPMW}. Indeed the difference between the super-modes' propagation constants ($\beta_{\rm s} - \beta_{\rm a} = \Delta_{\rm sa} \neq 0 $) gives rise to a coupling length $L_0$, defined at each frequency as half the beating length of the modes: $L_0=\pi/\Delta_{\rm sa}$\cite{burns1988}. At the same time anti crossing happens around the degeneration frequency with a bandwidth related to the coupling strength (\subref{imgCOUPLER}{a}). To confirm that the modes couples (whose dispersion is depicted in the main \subref{imgCMODES}{}) are symmetric-antisymmetric pairs we look at their even and odd recombinations and retrieve field profiles compatible with the uncoupled \glspl{TPHEM} (\subref{imgCMODES}{} inset).
\subsection*{Hybrid D/CD coupler}
\begin{figure}
\begin{center}
\noindent
	\includegraphics[width=\columnwidth]{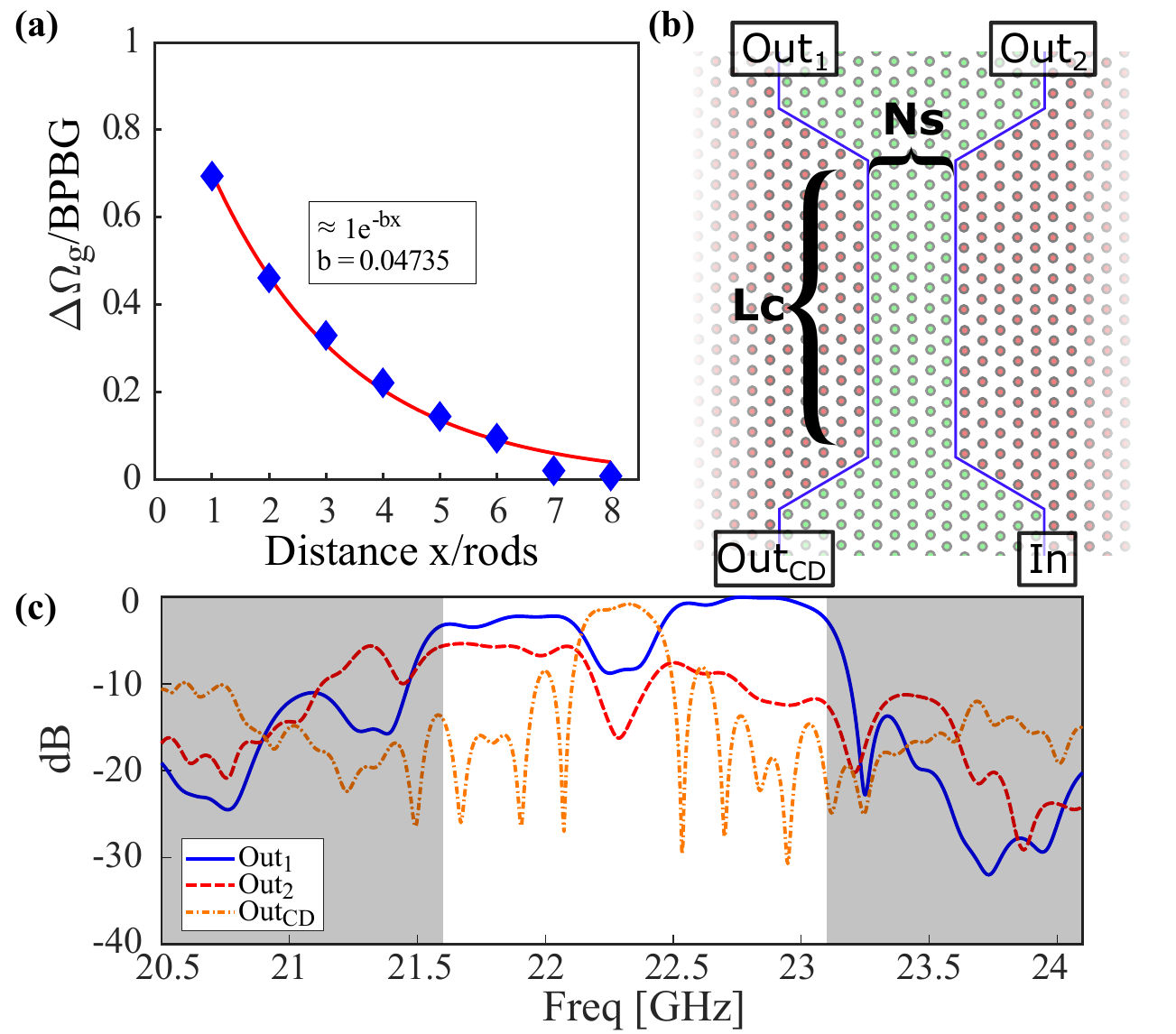}
  	\caption{(color online)
  	\textbf{(a)} Normalized bandwidth of the secondary frequency gap as a function of inter-guide separation. \glspl{TPMW}.
  	\textbf{(b)} Directional Coupler Structure: Red and greed dots represent rods with the air gap in opposite position. Blue lines indicates \glspl{TPMW}.
  	\textbf{(c)} Transmission spectra for a coupler with $L_c=30$ and $N_s=5$.
  	}
  	\label{imgCOUPLER}
\end{center}
\end{figure}
As a proof of concept for the behavior of the \gls{CTPMW} structure we now show an hybrid Directional/\gls{CD} coupler that acts on topological states.
One of the most critical points in conventional directional couplers is the design of the input/output tapering sections. In these regions two waveguides are bent in order to bring them close together down to a minimum distance in which the mutual interaction between the two is sufficiently strong. These bends normally introduce non-negligible losses that can only be addressed by increasing the curvature radius and, consequently, device sizes. A topological directional coupler is, on the contrary, immune to these losses and as such it provides a straightforward way to decrease the footprint of photonic devices that are based on a high number of directional couplers. Furthermore, the unique features of the \glspl{CTPMW} allow for the design of a device which is at the same time a Directional and a \acrlong{CD} coupler, depending on the input frequency, providing rich spectral features.\par
The basic structure of our topological directional coupler is illustrated in \subref{imgCOUPLER}{b} in which the input/output sections are clearly visible together with the interaction section. Our design depends on 2 parameters: The inter-waveguide separation $N_s$ that controls the relative bandwidth of the contra-directional region (see \subref{imgCOUPLER}{a}) and the interaction length $L_c$ that controls the splitting ratio in directional coupling regime. In our numerical simulations the uncoupled \glspl{TPMW} eigenmodes have been used as input and output, but every other kind of excitation can be used, including the previously discussed \gls{CWG} launcher.
\par
When exciting the input port for frequencies outside of the secondary gap, the propagating field on the input \gls{TPMW} overlaps with an even superposition of both \gls{CTPMW} super-modes $\psi_1 = 1/2(\psi_s + \psi_a)$. After propagating for a coupling length $L_0$ the two super-modes acquire a $\pi$ phase shift, producing a field completely localized at the second interface ($\psi_2 = 1/2(\psi_s - \psi_a)$). In this regime the device behaves as an optical directional coupler where the length of the interaction section defines the splitting ratio $\eta_s = P_{\mathrm{cross}}/(P_{\mathrm{cross}} + P_{\mathrm{bar}})$. For frequencies belonging to the anticrossing bandwidth, on the contrary, there are no allowed propagating states in the \glspl{CTPMW}. An input wave should be back-reflected towards the source but, as previously anticipated, this is forbidden by spin conservation and the only path that can be followed by the propagating wave is to couple to the phase matched \gls{Bw} mode of the coupled \gls{TPMW}. In this regime the coupling \acrlong{CD}, the transmittance peak of the \gls{CD} coupling rapidly approaches 1 with increasing coupling section length $L_c$, meaning that the coupling is fairly strong, while the bandwidth of the \gls{CD} effect is only influenced by the separation between the two waveguides and exponentially decreases for increasing distances. \par
\gls{CD} coupling is a well-known phenomenon in literature \cite{qiu2003, jandieri2014} and it is normally achieved by using Bragg gratings \cite{shi2013}. Interestingly enough however, because of the symmetries that defines spin states, \gls{CD} coupling appears as the predominant coupling phenomenon in QSH-like PTIs. Topological \gls{CD} coupling happens without the need of designing an appropriate Bragg grating and, if the coupling is strong enough, it can be present on a very large portion of the \gls{TPMW} operating bandwidth ( see \subref{imgCOUPLER}{a}). Moreover, spin orthogonality for the uncoupled \gls{TPMW} impairs self back-coupling, a phenomenon for which the input \gls{Fw} mode is coupled to a \gls{Bw} mode of the same waveguide rather than a \gls{Bw} mode of the coupled waveguide. In conventional Bragg-assisted \gls{CD} couplers this unwanted phenomenon is addressed by introducing a detuning between the two coupled waveguides, which in turn has detrimental effects in the co-directional coupling regime. Conversely, spin conservation allows for a perfectly balanced design which conserves the device functionality also in the Directional coupling regime.\par
A coupling section length of $30a_0$ and a guide separation of $N_s = 5$ rods produces a device with a $\approx 50\%$ splitting ratio for $f<22.1$GHz, a complete cross state for $f>22.5$GHz and almost unitary contra-directional coupling transmission for $f\in[22.1, 22.5]$GHz (\subref{imgCOUPLER}{c}). The described hybrid directional/\gls{CD} coupler is simulated through Full-Wave simulations on CST MWS with matched impedance boundary condition in order to eliminate reflections on the boundaries. To obtain the transmission diagram of \subref{imgCOUPLER}{c} we define 6 field probes along the \gls{TPMW} interface, close to each device port, with a relative distance of $a_0/2$. Then we extract the frequency dependent field intensity at each port by mediating the field intensities recorded by each of the 6 field probes in order to smooth the interference pattern as described in \cite{lai2016}. Finally the transmittance at any physical port is defined as the ratio of the port field intensity and the field intensity measured at the input port.
\section*{Conclusions}
We illustrated how load matching procedure can be applied to design a modal launcher for topological modes. Our optimized circular waveguide transition has a relative matching bandwidth of $\approx 73\%$ with respect to the operating bandwidth of the \acrlong{TPMW} and it can be used to directly probe topological protection, by observing broadband perfect transmission around a very sharp bend of the topological waveguide.\par
We studied the coupling mechanisms of two interacting \glspl{TPMW} by illustrating \textit{spin (inter modal)} coupling and \textit{inter-spin (modal)} coupling. These happen in distinct spectral regions and give rise to different coupling phenomena, respectively \acrlong{CD} and Directional coupling. Finally we presented a simple design for a device that implements a topological Directional-Contra Directional coupler. Our proposed design can be used to route a topological mode through three output ports and can be used to realize devices as beam splitters, interferometers and routers based entirely on topological propagation. Topological protection also makes the device less affected by the bends introduced by the tapered sections and keeps the design robust with respect to a class of fabrication defects (in particular missing or misplaced rods). The topological Directional Coupler might also find applications in testing Topological Photonics to the quantum regime since a $50\%$ beam splitter is often mandatory in many quantum optics experiments.

\end{document}